# Comment: Citation Statistics

**David Spiegelhalter and Harvey Goldstein**

We welcome this critique of simplistic one-dimensional measures of academic performance, in particular the naive use of impact factors and the h-index, and we can only extend sympathy to colleagues who are being judged using some of the techniques described in the paper. In particular we welcome the report's emphasis on the need for careful modeling of citation data rather than relying on simple summary statistics. Our own work on league tables adopts a modeling approach that seeks to understand the factors associated with institutional performance and at the same time to quantify the statistical uncertainty that surrounds institutional rankings or future predictions of performance. In the present commentary we extend this approach to an analysis of the 2008 UK Research Assessment Exercise (RAE) for Universities.

Before we describe our analysis it is important to comment on an important modeling problem that arises in the analysis of citation data, alluded to but not discussed in detail in the report, nor, as far as we know, elsewhere. A principal difficulty with indices such as the h-index or simple citation counts is that there are inevitable dependencies between individual scientists' values. This is because a citation is to a paper with, in general, several authors, rather than to each specific author. Thus, for example, if two authors nearly always write all their papers together, they will tend to have very similar values. If they belong to the same university department then their scores do not supply independent bits of information in compiling an overall score or rank for that department. Currently this issue is recognized in the RAE, albeit imperfectly, by the requirement that the same paper cannot be entered more than once by different authors for a given university department. In a citation based system this would also need to be recognized.

In addition, if our two authors were in different, competing departments, we would also need to recognize this since the dependency would affect the accuracy of any comparisons we make. We also note that this will, to some extent, affect our own analyses that we present below, and it will be expected to overestimate the accuracy of our rankings. Unfortunately we have no data that would allow us to estimate, even approximately, how important this is. To deal with this problem satisfactorily would involve a model that incorporated "effects" for each author and the detailed information about the authorship of each paper that was cited. Goldstein (2003, Chapter 12.5) describes a multilevel "multiple membership" model that can be used for this purpose, where individual authors become level 2 units and papers are level 1 units.

The UK Research Assessment Exercise was published on 18th December 2008, covering the years 2001–2008. 52,409 staff from 159 institutions were grouped into 67 "units of assessment" (UOA): up to 4 publications for each individual were considered as well as other activities and markers of esteem. Panels drawn from around 1000 peer reviewers then produced a "quality profile" for each group, summarizing in blocks of 5% the proportion of each submission judged by the panels to have met each of the following quality levels: "world-leading" (4*), "internationally excellent" (3*), "internationally recognized" (2*), "nationally recognized" (1*), and "unclassified." This procedure is notable in terms of its use of peer judgment rather than simple metrics, and allowing a distribution of performance rather than a single measure. All the data is available for downloading (Research Assessment Exercise, 2008).

*David Spiegelhalter is Winton Professor of Public Understanding of Risk, Statistical Laboratory, Centre for Mathematical Sciences, Wilberforce Road, Cambridge CB3 0WB, UK. Harvey Goldstein is Professor of Social Statistics, University of Bristol, 35 Berkeley Square, Bristol BS8 1JA, UK.*

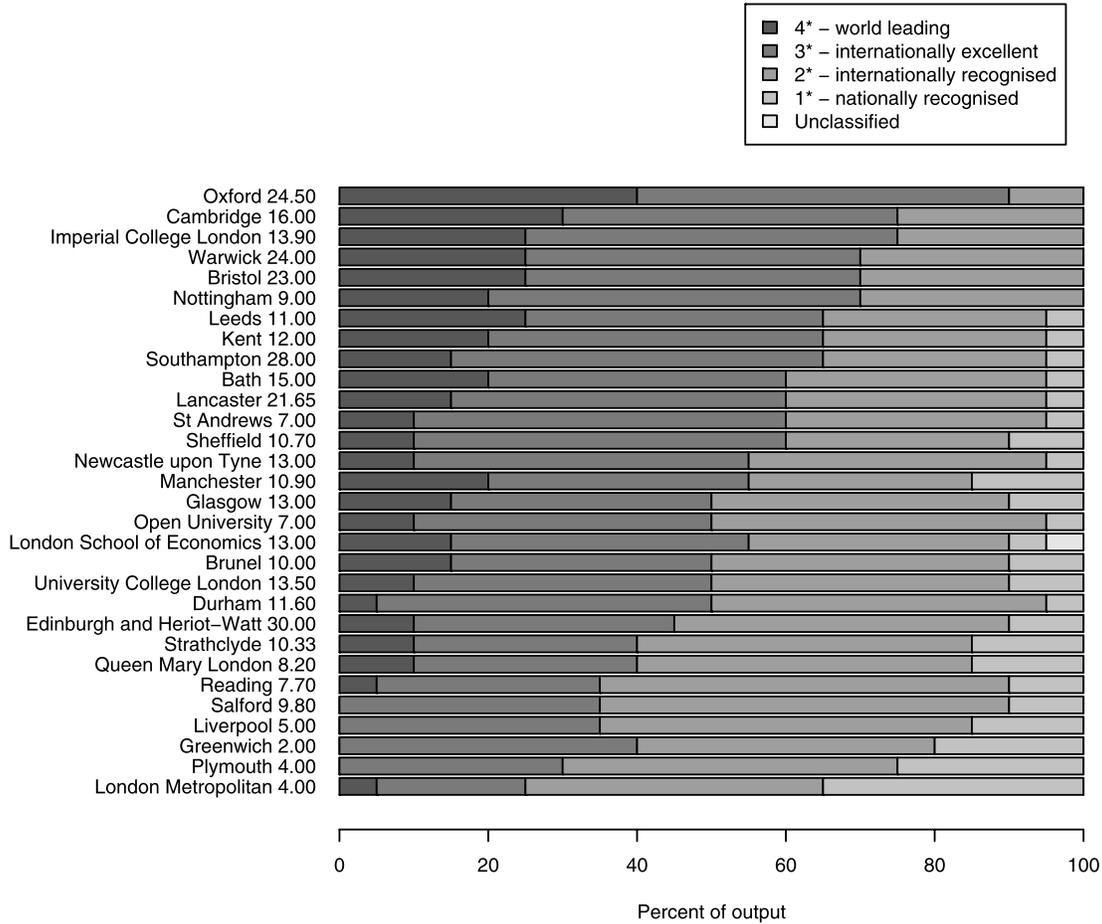

Fig. 1. *"Quality profiles" for 30 groups under UOA22 "Statistics and Operational research": UK Research Assessment Exercise 2008, ranked according to mean score: numbers of staff taken into account are shown.*

Figure 1 shows the results relevant for most statisticians: the 30 groups entered under UOA22: "Statistics and Operational Research." These have been ordered into a league table using the average number of stars which we shall term the "mean score," which is the procedure adopted by the media. Also reported is the number of full-time equivalent staff in the submission. Controversy surrounds this number as it is unknown how selective institutions were in submitting staff—it was originally intended that the total pool of staff would also be reported but late in the day there were objections raised as to the definitions of eligibility and this requirement was dropped.

The financial consequences of this whole exercise concern the distribution of around £1.5 billion of future funding. After publication of the quality profiles it was revealed that for funding purposes $4^*, 3^*, 2^*, 1^*$ outputs would be weighted proportional to $7, 3, 1, 0$: in further analysis we consider the "mean funding score" as $7p_4 + 3p_3 + p_2$, where $p_i$ is the proportion of outputs given $i$ stars.

In their report, Adler and colleagues argue that statistical analysis of performance data requires some concept of a model, and the provision of a quality profile rather than just a single number suggests it could be used for this purpose. We might first view the quality profile as representing the sampling distribution of material arising from each group, in fact a single Multinomial observation with probability $(p_4, p_3, p_2, p_1)$: if, in the spirit of a bootstrap, we simulate from these distributions and rank the institutions at each iteration, we can produce a distribution for the predicted rank of a random future output from each group as shown in Figure 2.

We note the substantial overlap of the distributions: in fact the rank distributions are highly multimodal due to the extreme number of ties at each iteration, which explains the somewhat anomalous results for some groups in which the median rank

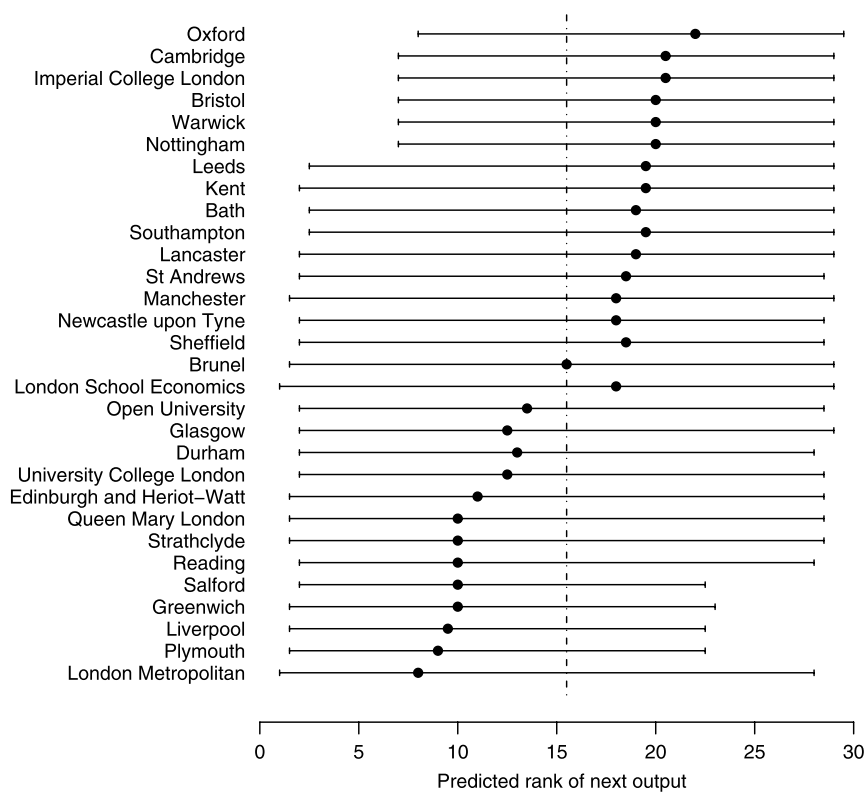

FIG. 2. *Predicted rank of a future output from each group: median and 95% intervals are shown based on 10,000 iterations.*

order is substantially different from the mean-score order in which the institutions are plotted.

We are not, however, particularly interested in a single output and instead we may want to focus on the accuracy with which a summary parameter, such as the underlying mean funding score, is known: we treat this as an illustration of a general technique for analyzing any summary measure arising from a specified weighting. It then seems reasonable to take into account the quantity of information underlying the quality profile: each individual contributes 4 publications and the publications count for 70% of the quality profile, and so we shall take a rough "effective sample size" as 6 outputs per staff member. Note that this does not mean that we are treating the publications as being a random sample from a larger population, but as relevant information connected through a probability model with some underlying parameter which may, in our particular illustration, be interpreted as the expected funding score of future outputs.

It would be possible to convert to ordered categorical data by multiplying the quality profile for each group by the number of publications taken into account (6 times the number of staff). Here, for the sake of simplicity, we have assumed a normal sampling distribution by estimating a standard error of the mean funding score as the square root of the sample variance of the profile divided by 6 times the number of staff.

Figure 3a shows the resulting estimates and 95% intervals for the mean scores. Treating these as normal distributions we can simulate future mean scores, rank at each iteration, and form a distribution for the "true" rank of each group. These are summarized in Figure 3b.

We see that for 14 out of 30 groups the 95% interval for the mean funding score overlaps the overall mean for all groups. Correspondingly we can identify 14 groups for which the 95% interval for their "true" rank, based on their mean funding scores, lies in either the top or bottom half. Both the mean funding scores and ranks, particularly for the smaller institutions, are associated with considerable uncertainty and this should warn against over-interpretation of either. If desired this could provide a basis for allocation into one of three groups for resource allocation purposes, although we would not necessarily recommend such a procedure.

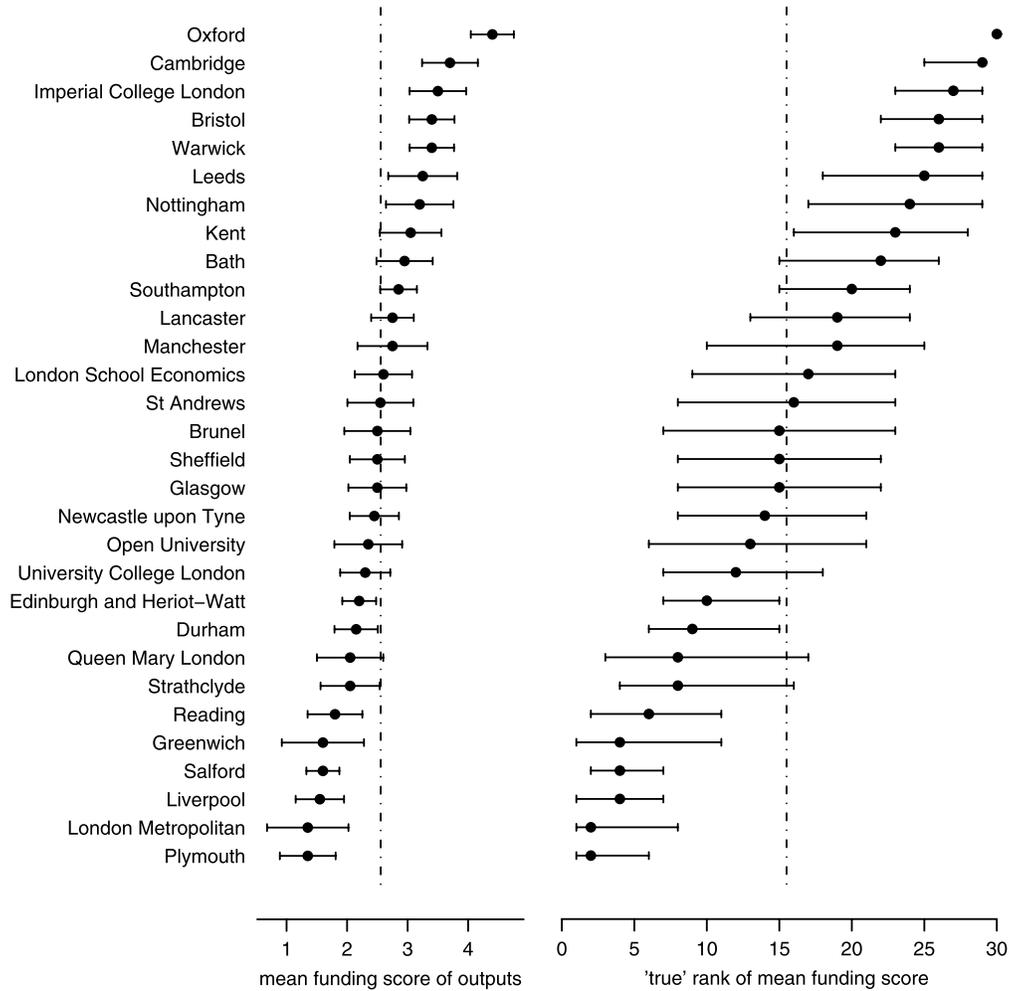

Fig. 3. *(Left) Estimates and intervals for expected funding score of outputs from each group. (Right) Summary of distribution of ranks of expected scores. Median and 95% intervals are shown based on 10,000 iterations.*

We could, in principle, take this analysis further by noting that if we are really interested in predicting future performance, then we should be taking into account the possibility of regression-to-the-mean, recognizing the variation within each institution that would be expected over time. We could do this by fitting a hierarchical/multilevel model where conditioning takes place on the current scores (see Goldstein and Leckie, 2008, for an example using school league tables). We could adjust for background factors such as available resources in order to reduce the within-institution variability and to help satisfy relevant exchangeability assumptions, and so produce an "adjusted" institution effect. Whether we use this adjusted effect, or the fitted mean, as a basis for comparing groups would depend on the purpose: if we were university administrators wanting to know whether a group had done well given the resources available, then we would examine the adjusted affect. If, however, we wished to use the current scores simply to allocate income, then the fitted mean would be appropriate: see Goldstein and Leckie (2008) for a close examination of the potential role for different kinds of adjustments when comparing schools. In practice it is likely that such an analysis would be considered too complex.

In conclusion, we agree with the Report's strictures on the meaning of citation counts and would go further and argue that citations form a rather bizarre measure of research performance, as if the sole purpose of research was to provide material for other researchers. If they are to be used, we would argue that they be analyzed within a statistical modeling framework that fully incorporates uncertainty and dependency. As we have shown, for example, in Figure 3b, this could help to guide fund-ing decisions by avoiding fine distinctions that may reflect little more than random noise. But citations alone, no matter how carefully analyzed, can only provide one measure of performance, and we feel strongly that they should be part of a broader profile that takes into account other measures of real world impact and is assessed using peer judgement rather than mechanistic and spuriously "objective" processes.